\documentclass[8.5pt,twoside,twocolumn]{article}
\usepackage[super,sort&compress,comma]{natbib} 
\usepackage{mhchem,amsmath,amssymb}
\usepackage{mciteplus}
\mciteSetMaxCount{main}{bibitem}{47}
\usepackage{times}
\usepackage{sectsty}
\usepackage{balance} 

\usepackage{graphicx,amsmath} 
\usepackage{lastpage}
\usepackage[format=plain,singlelinecheck=false,font=small,labelfont=bf,labelsep=space]{caption} 
\usepackage{fancyhdr}
\begin{document}

\thispagestyle{plain}
\fancypagestyle{plain}{
\fancyhead[L]{\includegraphics[height=8pt]{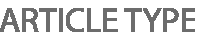}}
\fancyhead[C]{\hspace{-1cm}\includegraphics[height=20pt]{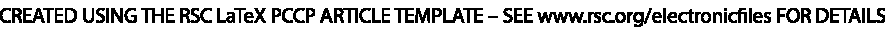}}
\fancyhead[R]{\includegraphics[height=10pt]{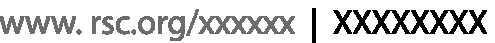}\vspace{-0.2cm}}
\renewcommand{\headrulewidth}{1pt}}
\renewcommand{\thefootnote}{\fnsymbol{footnote}}
\renewcommand\footnoterule{\vspace*{1pt}%
\hrule width 3.4in height 0.4pt \vspace*{5pt}} 
\setcounter{secnumdepth}{5}

\makeatletter 
\def\subsubsection{\@startsection{subsubsection}{3}{10pt}{-1.25ex plus -1ex minus -.1ex}{0ex plus 0ex}{\normalsize\bf}} 
\def\paragraph{\@startsection{paragraph}{4}{10pt}{-1.25ex plus -1ex minus -.1ex}{0ex plus 0ex}{\normalsize\textit}} 
\renewcommand\@biblabel[1]{#1}            
\renewcommand\@makefntext[1]%
{\noindent\makebox[0pt][r]{\@thefnmark\,}#1}
\makeatother 
\renewcommand{\figurename}{\small{Fig.}~}
\sectionfont{\large}
\subsectionfont{\normalsize}

\fancyfoot{}
\fancyfoot[LO,RE]{\vspace{-7pt}\includegraphics[height=9pt]{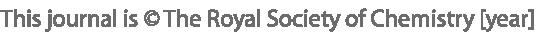}}
\fancyfoot[CO]{\vspace{-7.2pt}\hspace{12.2cm}\includegraphics{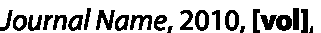}}
\fancyfoot[CE]{\vspace{-7.5pt}\hspace{-13.5cm}\includegraphics{RF.eps}}
\fancyfoot[RO]{\footnotesize{\sffamily{1--\pageref{LastPage} ~\textbar  \hspace{2pt}\thepage}}}
\fancyfoot[LE]{\footnotesize{\sffamily{\thepage~\textbar\hspace{3.45cm} 1--\pageref{LastPage}}}}
\fancyhead{}
\renewcommand{\headrulewidth}{1pt} 
\renewcommand{\footrulewidth}{1pt}
\setlength{\arrayrulewidth}{1pt}
\setlength{\columnsep}{6.5mm}
\setlength\bibsep{1pt}

\twocolumn[
  \begin{@twocolumnfalse}
\noindent\LARGE{\textbf{Glassy dynamics of crystallite formation: The role of covalent bonds}}
\vspace{0.6cm}

\noindent\large{\textbf{Robert S. Hoy\textit{$^{a,b,\ast}$} and Corey S. O'Hern\textit{$^{a,b}$}}}\vspace{0.5cm}

\noindent\textit{\small{\textbf{Received Xth XXXXXXXXXX 20XX, Accepted Xth XXXXXXXXX 20XX\newline
First published on the web Xth XXXXXXXXXX 200X}}}

\noindent \textbf{\small{DOI: 10.1039/b000000x}}
\vspace{0.6cm}

\noindent \normalsize{We examine nonequilibrium features of collapse behavior in model polymers with competing crystallization and glass transitions using extensive molecular dynamics simulations.  By comparing to ``colloidal'' systems with no covalent bonds but the same non-bonded interactions, we find three principal results: \textbf{(i)} Tangent-sphere polymers and colloids, in the equilibrium-crystallite phase, have nearly identical static properties when the temperature $T$ is scaled by the
crystallization temperature $T_{\rm cryst}$; \textbf{(ii)} Qualitative features of nonequilibrium relaxation below $T_{\rm cryst}$, measured by the evolution of local structural properties (such as the number of contacts) toward equilibrium crystallites, are the same for polymers and colloids; and \textbf{(iii)} Significant quantitative differences in rearrangements in polymeric and colloidal crystallites, in both far-from equilibrium and near-equilibrium systems, can be understood in terms of chain connectivity.  These results have important implications for understanding slow relaxation processes in collapsed polymers, partially folded, misfolded, and intrinsically disordered proteins.
}
\vspace{0.5cm}
 \end{@twocolumnfalse}]

\footnotetext{\textit{$^{a}$~Department of Mechanical Engineering and Materials Science, Yale University, New Haven, CT, USA 06520-8286}}
\footnotetext{\textit{$^{b}$Department of Physics, Yale University, New Haven, CT 06520-8120}}
\footnotetext{\textit{$^{\ast}$~Email: robert.hoy@yale.edu}}

\section{Introduction}

Collapse transitions of single chain polymers induced by changing
control parameters such as temperature or solvent quality yield rich
nonequilibrium behavior when the rate at which these control
parameters are changed exceeds characteristic (slow) dynamical rates.
Investigating the glassy dynamics of polymer collapse is important for
understanding \textit{e.g.}\ crystallization kinetics and protein misfolding,
yet the majority of studies have focused on equilibrium behavior.
In this manuscript, we characterize the nonequilibrium and
near-equilibrium collapse and crystallization dynamics of single
flexible polymer chains.

We employ a minimal model that yields competing crystallization and glass
transitions.
Monomers are modeled as monodisperse tangent spheres with hard-core-like repulsive and short-range attractive interactions.
Recent studies\cite{taylor09,seaton10} have shown that in
equilibrium, model polymers with narrow square-well interactions
exhibit direct ``all-or-nothing'' crystallization transitions that
mimic the discrete folding transition observed in experimental studies
of proteins.\cite{proteincites}
Short-range attractions also give rise to degenerate, competing ground states, which
kinetically hinder collapse to equilibrium crystallites.
The associated rugged energy landscapes are believed
to control the behavior of intrinsically disordered
proteins.\cite{socci94,foffi00,cellmer07}

A novel aspect of our work is quantitative comparison of polymer collapse dynamics to that of
`colloidal' systems with the same secondary interactions but no
covalent bonds.  
Polymers are distinguished from other systems by their topology;
connectivity and uncrossability constraints imposed by covalent
backbone bonds give rise to cooperative dynamics~\cite{doi86} and
phase transitions~\cite{dimarzio06} not present in nonpolymeric
materials. 
Our choice of tangent monodisperse spheres yields identical low-energy
states for polymers and colloids, but (because of the covalent
backbone) very different free energy landscapes.\cite{footprevcollpol}  This greatly
facilitates a robust comparison of crystallite formation and growth
dynamics that isolates the role of topology and allows us to isolate and quantify the contributions of the covalent bonds and chain uncrossability to cooperative rearrangements and slow dynamics during collapse.
We perform this comparison using extensive molecular dynamics simulations of
thermal-quench-rate-dependent collapse and post-quench growth of
polymeric and colloidal crystallites.   
 
\begin{figure}[htbp]
\centering
\includegraphics[width=3.25in]{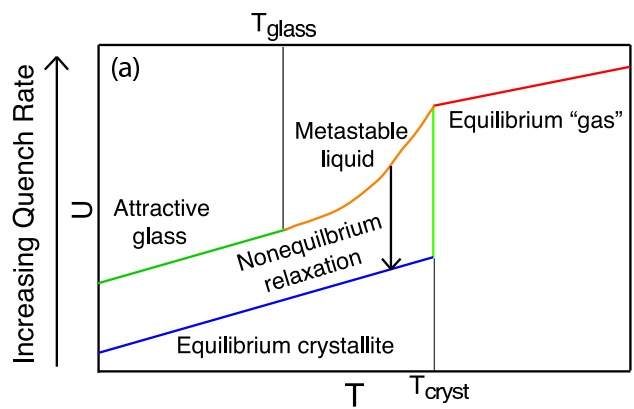}
\caption{Schematic of potential energy versus temperature $T$ for systems with short-range attractive interactions cooled at varying quench rates $\dot{T}$.\cite{sastry98}  We compare the nonequilibrium and near-equilibrium crystallization dynamics of polymers and colloids by quenching systems at various rates and observing relaxation dynamics below $T_{\rm cryst}$.}
\label{fig:schematic}
\end{figure}

Figure \ref{fig:schematic} depicts rate-dependent collapse behavior of systems interacting via hard-core-like repulsions and short-range attractions.\cite{sastry98}
In the limit of slow quench rates $|\dot{T}| < |\dot{T}^*|$, where $\dot{T}^*$ is a critical quench rate, finite systems exhibit a first-order-like transition from a high-temperature ``gas'' (for polymers, a self-avoiding random coil) phase to crystallites.
The equilibrium transition occurs if $|\dot{T}|$ is small compared to key relaxation rates,
such as the crystal nucleation rate $r$ and rate $s$ of large rearrangements in compact structures.  
At larger $|\dot{T}|$, systems fall out of equilibrium, and pass onto the metastable liquid
branch. 
If $T$ becomes low enough such that $s(T) \gg |\dot{T}|$, the systems become glassy and disorder is frozen in at $T = T_{\rm glass}$.
Otherwise, systems relax towards equilibrium crystallites, as indicated by the downward arrow.

Our analyses compare collapse behavior within the framework of Fig.\ \ref{fig:schematic}.
We find that polymers and colloids behave similarly in many ways, and differences can be linked directly to chain topology.
Near-equilibrium crystallites possess similar static structure when $T$ is scaled by
the equilibrium crystallization temperature $T_{\rm cryst}$.
Relaxation dynamics at fixed $T/T_{\rm cryst}$ are qualitatively
similar, but quantitatively quite different for polymers and colloids.
Using multiple measures of crystalline order and particle rearrangements, we
quantify how restrictions on local motion imposed by the covalent
backbone slow the approach to equilibrium by eliminating `monomeric'
relaxation mechanisms.  Rearrangements in polymeric crystallites are
required to be more cooperative, and hence are slower, than in their
colloidal counterparts.  
We validate these results by examining whether these trends vary significantly with system size (\textit{i.e.} chain length) $N$.
For $N$ varying over a range typically studied in single chain polymer crystallization experiments,\cite{mandelkern90} increasing system size does not affect qualitative trends, but quantitatively strengthens them.

The outline of the rest of the paper is as follows.  In Section
\ref{sec:methods}, we describe our model, simulation protocol, and
metrics used to analyze our data.  Section \ref{sec:results} presents
results for thermal quenches over a wide range of $|\dot{T}|$,
nonequilibrium evolution of crystallite properties at fixed $T_f <
T_{\rm cryst}$, and a detailed comparison of rearrangements in polymers
and colloids.  Finally, in Section \ref{sec:discussion}, we summarize
our findings and place them in context of other recent experimental
and simulation studies.

\section{Methods}
\label{sec:methods}

Recent Monte Carlo simulations have employed advanced sampling
techniques, such as topology-changing bridging moves, to investigate
the equilibrium phase behavior\cite{taylor09,seaton10} of
single-chain polymers. In contrast, molecular dynamics (MD)
simulations with physically realistic dynamics are better able to capture the
complex, coordinated rearrangement events associated with the glassy
dynamics of crystallization.  

\begin{figure}[htbp]
\includegraphics[width=3.1in]{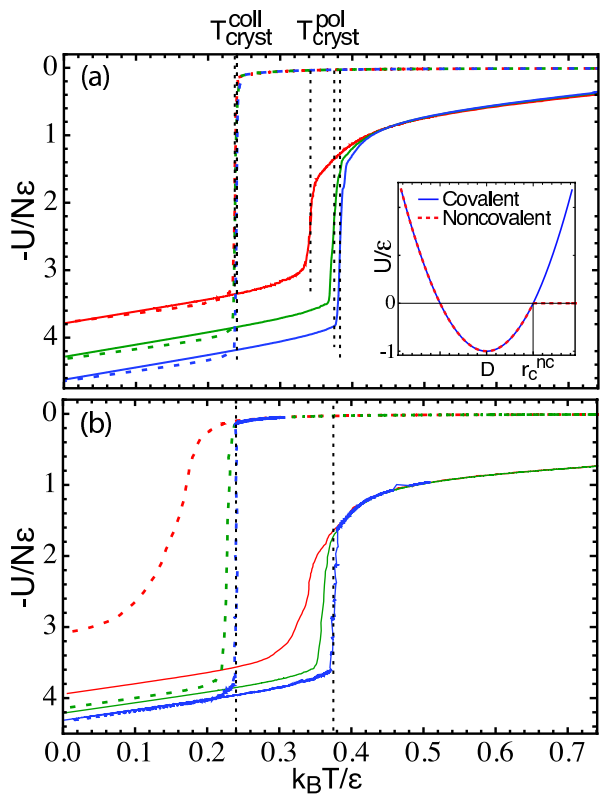}
\caption{(a) System size dependence and (b) quench rate dependence of potential energy versus temperature from molecular dynamics
simulations of polymers (solid curves) and colloids (dashed curves). 
Vertical dashed lines indicate $T_{\rm cryst}$.  
Panel (a) shows results for a single slow ($|\dot{T}| = 2.5\cdot10^{-8}$) quench rate for $N = 40$ (red), $N=100$ (green), and $N = 250$ (blue).  
Panel (b) shows results for $N=100$ systems.   Colors indicate $|\dot{T}| = 10^{-4}$ (red), $10^{-6}$ (green) and $10^{-8}$ (blue).  Results in (a) are averaged over 40 statistically independent samples, while results in (b) are averaged over $104$ statistically independent samples for the higher two quench rates and $8$ for the lowest.  The inset to (a) illustrates the interaction potential (Eq.~\ref{eq:perturbed}) employed in our simulations.}
\label{fig:one}
\end{figure}

In our studies, both colloidal and polymeric systems consist of $N$ identical
spherical monomers that interact via the harmonic ``sticky-sphere''
potential shown in Fig.\ \ref{fig:one}(a):
\begin{equation} 
U_{\rm harm}(r) = \Bigg{\{}\begin{array}{ccc}
-\epsilon + \displaystyle\frac{k}{2}\left(\displaystyle\frac{r}{D}-1\right)^{2} & , & r < r_c\\
& & \\
\normalsize 0 & , & r > r_c
\end{array},
\label{eq:perturbed}
\end{equation}
where $\epsilon$ is the intermonomer binding energy, $D$ is the
monomer diameter, and $k = 1600\epsilon$ is the spring constant.
The only difference between colloidal and polymeric interactions (inset to Fig.\ \ref{fig:one}(a))
is that in polymeric systems, the monomers are linked into a linear
chain connected by $N-1$ permanent covalent bonds.\cite{footnobondcross}
Different values for $r_c$ are used for covalently and noncovalently bonded monomers: $r^c_c/D = \infty$ and $r^{nc}_c/D = 1 +\sqrt{2\epsilon/k}$, respectively.  

Newton's equations of motion are integrated using the velocity-Verlet
method with a timestep $dt = \tau/800$, where $m$ is the monomer mass
and $\tau=\sqrt{mD^2/\epsilon}$.  We determined that this timestep was
sufficiently small by examining $dt$-dependence of the velocity
autocorrelation function $v_{ac}(t)$ in simulations at high
temperature; no statistically significant dependence was found for $dt
\leq \tau/600$.  Below, we express length scales, energies,
times, rates and temperatures in units of $D$, $\epsilon$, $\tau$,
$\tau^{-1}$, and $\epsilon/k_B$, respectively.  The temperature $T$ is
controlled via a Langevin thermostat with damping time $10$.  A
periodic cubic simulation cell with volume $L^{3}$ fixes the monomer
density $\rho$.  We present results for $\rho = 0.01$, which is in the
dilute limit.

Systems are initialized in random walk initial conflgurations (for
polymers) and random nonoverlapping initial positions (for colloidal
systems).  They are then equilibrated at high temperature, $k_B
T_i/\epsilon=0.75$, for times long compared to the time over which the
self intermediate scattering function decays to zero.  At $T = T_i$,
polymeric systems are in the good-solvent (self-avoiding coil) limit,
and colloidal systems are in an ideal-gas-like state.  Following
equilibration, systems are thermally quenched at various rates $\dot{T}$.  In
our ensemble-averaging approach, states from which thermal quenches
are initiated are separated by times long compared to structural
relaxation times, and thus are statistically independent.

The quenches are either continued to $T=0$ or terminated at $T = T_f <
T_{\rm cryst}$.  In the latter case, we continue the runs at fixed
$T_f$.  We choose $T_f/T_{\rm cryst} = 7/8$ to suppress finite-$N$
fluctuation effects associated with thermal broadening of the phase
transition to crystallites, \cite{chakrabarty10} {\it i.e.}
$T_f/T_{\rm cryst} \lesssim 1-N^{-1/2}$, yet allow sufficiently fast
relaxation to be captured within the limits of available computational
resources for the $N$ considered here (40, 100, and 250).  As we will
show below, this procedure yields particularly interesting results for
nonequilibrium relaxation following quenches at moderate $|\dot{T}|$.
$T_f/T_{\rm cryst} = 7/8$ is also comparable to temperatures used in many experimental and simulation studies of crystallite nucleation and growth in supercooled colloidal and polymeric systems.\cite{tenwolde96,dalnoki}

We will examine several order parameters to characterize nucleation,
growth and rearrangements of crystallites as a function of temperature
and time following thermal quenches to $T_f/T_{\rm cryst} < 1$.  These
order parameters are generated from the adjacency matrix $\bar{A}$;
$A_{ij} = A_{ji} = 1$ when particles $i$ and $j$ are in contact, {\it
i.e.}  when the position vectors satisfy $|\vec{r}_{i} - \vec{r}_{j}|
< r^{nc}_c$, and $0$ otherwise.

Our interaction potential (Eq.\ \ref{eq:perturbed}) promotes
contact-dominated crystallization dynamics.\cite{footnoicos}  For the large
$k/\epsilon$ and small $r_c$ employed here, the ground states for
colloidal and polymeric systems~\cite{biedl01} are simply the states
that maximize the number of pair contacts $N_c = \Sigma_{j>i} A_{ij}$
for a given $N$.\cite{arkus09, hoy10,footstickysphere} 
Further, our systems form crystallites possessing close-packed cores that increase
in size as $T$ decreases or equilibrium is
approached.\cite{footnoicos} We therefore measure the number of close
packed monomers $N_{cp} = \Sigma_{i=1}^{N} \delta(\Sigma_{j} A_{ij} - 12)$
and degree distribution $P(\eta)$, \textit{i.e.}\ the probability
for a particle to have $\eta$ contacts.  Note that $\sum_{i = 0}^{12}
iP(i) = 2N_c/N$ and $P(12) = N_{cp}/N$.  $P(\eta)$ contains more
information than $N_c$ and $N_{cp}$ since it describes the high-$\eta$
cores of crystallites as well as their surfaces, where monomers
naturally have lower $\eta$.\footnote{$P(\eta)$ is also closely related to the spectrum of eigenvalues of $\bar{A}$.\cite{chung96}  While these eigenspectra provide additional information on crystallite structure, we leave their examination for future studies of equilibrium systems.}  We will argue below that
the combination of $N_c$, $N_{cp}$ and $P(\eta)$ forms an effective
set of ``crystal-agnostic'' order parameters (in the spirit of Rein
ten Wolde \textit{et.al.}\cite{tenwolde96}).

We will also present results for the adjacency matrix autocorrelation
function $P_{\rm AMAC}(t_w,t')$ as a function of $t'$ after waiting
various times $t_w$ following thermal quenches to $T_f$:
\begin{equation}
P_{\rm AMAC}(t_w,t') = \left<\displaystyle\frac{\sum_{i,j > i}A_{ij}(t_w)A_{ij}(t_w+t')}{\sum_{i,j > i}A_{ij}(t_w)A_{ij}(t_w)}\right>,
\label{eq:PAMACdef}
\end{equation}
where the brackets indicate an ensemble average over independently
prepared samples, and the total time elapsed after termination of the quench is $t = t_w + t'$. 
For polymers, we exclude the contributions of
covalent bonds to $P_{\rm AMAC}$ by summing over $j > i+1$ rather than
$j > i$ in Eq.\ \ref{eq:PAMACdef}.  Both $P_{\rm AMAC}(t')$ and the
intermediate scattering function $S(q,t')$ evaluated at $qD \simeq
2\pi$ identify rearrangements of contacting neighbors, which control
the slow relaxation processes in colloidal systems with competing
crystallization and glass transitions.\cite{kob00,puertas02}
To capture the glassy dynamics, we examine systems with $t_w$ ranging over several orders of magnitude and $t' \gg t_w$.

\section{Results}
\label{sec:results}

In this section, we compare the collapse and ordering dynamics 
of colloids and polymers 
using two protocols.
Protocol (1) consists of decreasing the temperature from an
initially high value $T_i$ to zero using a wide range of thermal quench rates ${\dot T}$.
To analyze changes in structure with decreasing $T$, we measure the potential energy, $N_c$, and $N_{cp}$ over the full-range of temperature for each quench rate.  
Protocol (2) consists of quenching systems from $T_i$ to $T_f = (7/8)T_{\rm cryst}$
using a range of $\dot{T}$ and then monitoring structural order and rearrangement events within 
crystallites at $T = T_f$ as systems evolve toward equilibrium.
We measure $N_c(t)$, $N_{cp}(t)$, $P(\eta; t)$ and $P_{AMAC}(t_w,t')$,
to quantify evolution to more ordered states,
and also perform detailed studies of crystallite rearrangements in a ``pre-terminal'' relaxation regime where the systems slowly approach equilibrium.

\subsection{Protocol 1: Thermally Quench from High to Zero $T$}
        
{\it Potential Energy:} Figure \ref{fig:one} shows results for the scaled
potential energy $U/N\epsilon$ for colloidal and polymer systems quenched from
$T=T_i$ to zero.
Panel (a) shows results at low $|\dot{T}|$ for system sizes ranging over a factor of six in $N$, while panel (b) shows results for $N=100$ over a range of $|\dot{T}|$ spanning four orders of
magnitude.  
All results are consistent with the general picture of Fig.\ \ref{fig:schematic}, and illustrate both features common to polymers and colloids as well as differences arising from the presence of a covalent backbone.

At the lowest $|\dot{T}|$ considered ($\sim 10^{-8}$), both colloids and
polymers show sharp, first-order-like\cite{taylor09}
transitions at corresponding $T = T_{\rm cryst}$.  
Because of the narrowness of the attractive range of the potential well,\cite{hagen94, taylor09} no intermediate liquid state ({\it i.e.}, globules in the case of polymers) appears \cite{footnoicos, foot12}.  
In both cases, as in Fig.\ \ref{fig:schematic}, the equilibrium transitions are from gas-like
states to crystallites.
No significant quench rate dependence is observable
for $T > T_{\rm cryst}$, which indicates that all $|\dot{T}|$ are
sufficiently low to be near-equilibrium in this high temperature
regime.
For $|\dot{T}| \sim 10^{-8}$, polymers and colloids have the same energy at low $T$ to within
statistical noise, showing that this quench rate is slow enough to
be in the near-equilibrium limit for polymers ({\it i.e.}
$|\dot{T}^{*}| \gtrsim 10^{-8}$ for these systems.)
The $N$-dependence (panel (a)) shows only quantitative rather than qualitative differences.  
As $N$ increases, values of $U/N\epsilon$ in the $T \to 0$ limit decrease because larger crystallites possess more interior monomers.

Compared to colloids, polymers have lower $U/N\epsilon$ for $T > T_{\rm cryst}$ because of the permanent covalent bonds, which contribute $\sim k_B T/\epsilon - 1$.\cite{foot12}  
They also have higher absolute values of $T_{\rm cryst}$ (Table \ref{tab:Tcrysttab}). 
To zeroth order, $T_{\rm cryst}^{pol}/T_{\rm cryst}^{coll} \simeq 3/2$, a ratio which can be explained by a simple degree-of-freedom counting argument: while colloids have $3N-6$ nontrivial degrees of freedom, the stiff covalent bonds in polymers act like holonomic constraints, reducing the effective dimensionality of their phase space to $2N-5$ so that $T_{\rm cryst}^{pol} = [(3N-6)/(2N-5)]T_{\rm cryst}^{coll} \simeq 3/2$.\cite{taylor03}
This observation helps motivate our (protocol 2) studies comparing relaxation of polymeric and colloidal crystallites at equal values of $T/T_{\rm cryst}$ in terms of their different free energy landscapes.

\begin{table}[h]
\centering
\small\caption{Dependence of $T_{\rm cryst}$ on $N$ and topology.  Our data are consistent with detailed analyses \cite{dimarzio06} predicting $O(N^{-1/3})$ finite-size corrections to $T_{\rm cryst}^{pol}$.}
\begin{tabular}{lccc}
\hline
$N$ & $T_{\rm cryst}^{coll}$ & $T_{\rm cryst}^{pol}$ & $T_{\rm cryst}^{pol}/T_{\rm cryst}^{coll}$\\
\hline
40 & 0.237 & 0.342 & 1.44\\
100 & 0.240 & 0.375 & 1.56\\
250 & 0.239 & 0.383 & 1.60\\
\hline
\end{tabular}
\label{tab:Tcrysttab}
\end{table} 

Below $T_{\rm cryst}$, dramatic quench rate dependence appears.  Consistent with the general picture
of Fig.\ \ref{fig:schematic}, with increasing $|\dot{T}|$, the $T$-dependent potential
energy increases relative to the equilibrium-crystal limit.
Figure \ref{fig:one}(b) shows results for $N=100$.  Similar trends are observed for the other $N$ considered, with the main difference being that $|\dot{T}^{*}|$ increases and rate effects for $|\dot{T}| > |\dot{T}^{*}|$ become more dramatic with increasing $N$.

\begin{figure}[htbp]
\centering
\includegraphics[width=3.1in]{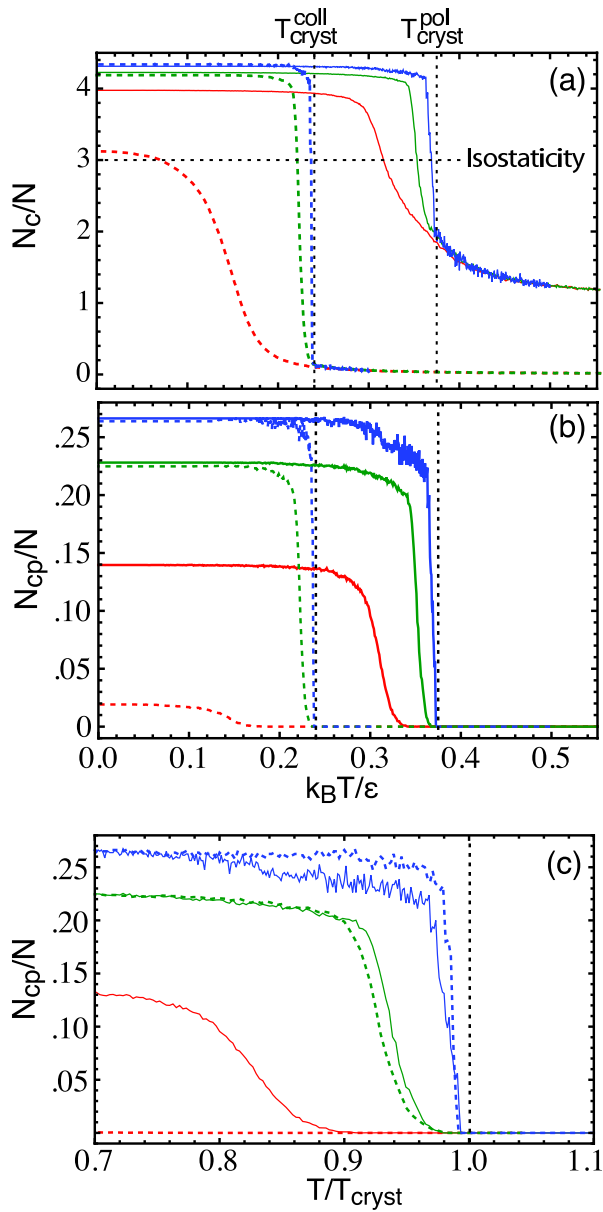}
\caption{Measures of local structural order, (a) $N_c/N$ and (b) $N_{cp}/N$, as a function of temperature
during thermal quenches from $T_i = 0.75$ to $T_f = 0$.  (c) Number of
close-packed spheres $N_{cp}/N$ plotted versus temperature reduced by
the crystallization temperature, $T/T_{\rm cryst}$.  The line color, dashing
scheme, and ($N = 100$) systems studied are the same as those in Fig.\
\ref{fig:one}(b).  Vertical dashed lines indicate $T_{\rm cryst}$ for colloids and polymers, and
the horizontal dashed line in panel (a) indicates the isostaticity
\cite{alexander98} threshold ($N_c/N = 3$).  We define $T_{\rm glass}$ as the temperature at which rearrangements cease during a constant rate quench and the slope $\partial N_c/\partial T$ becomes close to zero.}
\label{fig:NcNcpinquenches}
\end{figure}

{\it Contact number:}  We now discuss the temperature dependence of
local measures of structure during quenches. 
Figure \ref{fig:NcNcpinquenches} shows $N_c(T)$ and
$N_{cp}(T)$ plotted against $T$ and $T/T_{\rm cryst}$ for the same systems
analyzed in Fig.\ \ref{fig:one}(b).  $N_c/N$ and $N_{cp}/N$ are
particularly sensitive measures of crystallite equilibration.  They
increase monotonically with decreasing $|\dot{T}|$, but the variations
with rate are different for polymers and colloids.  For fast quenches,
polymers are more efficient crystal formers due to topological
connectivity and the resultant cooperative dynamics.\cite{doi86}  
An example of a cooperative polymer collapse mechanism not found in colloids is ``chain pull-in''; a nucleus forms between chemically nearby monomers, and covalent bonds pull in segments that are chemically adjacent to the nucleus.\cite{dill97} 
However, as exemplified by the $|\dot{T}| = 10^{-4}$ results,
this mechanism produces only small ordered cores with highly disordered
exteriors.   Note that for the fastest quench rates, the collapsed
structures possess $N_c/N \ge 3$, which is the minimal number of
contacts for mechanical stability\cite{alexander98} and corresponds to the isostaticity threshold shown in panel (a).

In Fig.~\ref{fig:NcNcpinquenches}(a), we show that $|\partial
N_c/\partial T| \to 0$ at a noticeably higher $T$ for polymers than
colloids, which indicates that large structural rearrangements cease,
and polymers possess a higher $T_{\rm glass}$ than colloids.\cite{kob00}  
Polymer chain backbones follow tortuous paths through
the interior of collapsed states \cite{hoy10, karayiannis09prl} and
chain uncrossability suppresses large rearrangements, so a higher
$T_{\rm glass}$ is consistent with the expected slower relaxation
dynamics for polymer crystallites.  

We find that measures of energy and contact number ($U$, $N_c$, and $N_{cp}$) plotted against the scaled temperature $T/T_{\rm cryst}$ for polymers and colloids all increasingly collapse as $|\dot{T}|$ decreases. 
Figure \ref{fig:NcNcpinquenches}(c) shows $N_{cp}(T/T_{\rm cryst})$ for different quench rates.
Over a broad range of $T$, the collapse of ensemble-averaged values of $N_c(T/T_{\rm cryst})$ and $N_{cp}(T/T_{\rm cryst})$ for $N = 40$ and $N = 250$ is of similar quality.\footnote{Collapse worsens only slightly with increasing $N$, indicating that $|\dot{T}^{*}|$ increases slowly with $N$.  Precise calculation of $|\dot{T}^{*}(N)|$ is outside the scope of this study, but more quantitative results for the $N$-dependence of characteristic relaxation times is given in Section \ref{subsec:protocol2}.} 
This is a nontrivial result for the small crystallites considered here, where ordering and surface-to-volume ratio changes with $N$ and finite-size effects are in principle important.
As shown in Table \ref{tab:groundstateorder}, terminal values of $\left<N_c/N\right>$ increase from $\sim 3.8$ to $\sim 4.7$ as $N$ increases from 40 to 250,\footnote{This variation is significant when one considers that the isostatic value $\left<N_c/N\right> = 3$ can be attained by systems as small as $N=16$,\cite{arkus11} while the limiting value for infinite $N$ (corresponding to defect-free close-packed crystals) is $\left<N_c/N\right> = 6$.} and values of $\left<N_{cp}/N\right>$ at $T=0$ increase even more (from $\sim0.15$ to $\sim0.35$) as the volume-to-surface-area ratio of crystallites increases and they form larger close-packed cores.
The collapse of $U(T/T_{\rm cryst})$, $N_c(T/T_{\rm cryst})$, and $N_{cp}(T/T_{\rm cryst})$ despite these $N$-dependent changes in order suggests that polymer and colloid crystallites, in equilibrium, occupy similar positions on their respective energy landscapes at equal values of
$T/T_{\rm cryst}$ even though the absolute $T$ are different. 

\begin{table}[h]
\centering
\small\caption{Variation with $N$ of $\left<N_c\right>$ and $\left<N_{cp}\right>$ during $|\dot{T}|=2.5\cdot10^{-8}$ quenches.  Results are from the same systems depicted in Fig.\ \ref{fig:one}(a), and values of $T_{\rm cryst}$ are given in Table \ref{tab:Tcrysttab}.  Results for $\left<N_c\right>/N$ are consistent with leading order $O(N^{-1/3})$ corrections away from the $N\to\infty$ value (6).}
\begin{tabular}{lccccc}
\hline
$N$ & $T$ & $\left<N_c^{coll}/N\right>$ & $\left<N_c^{pol}/N\right>$ & $\left<N_{cp}^{coll}/N\right>$ & $\left<N_{cp}^{pol}/N\right>$\\
\hline
40 & $\frac{7}{8}T_{\rm cryst}$ & 3.73 & 3.68 & 0.143 & 0.130\\
40 & 0 & 3.78 & 3.76 & 0.150 & 0.146\\
\hline
100 & $\frac{7}{8}T_{\rm cryst}$ & 4.31 & 4.24 & 0.259 & 0.239\\
100 & 0 & 4.33 & 4.29 & 0.265 & 0.255\\
\hline
250 & $\frac{7}{8}T_{\rm cryst}$ & 4.67 & 4.60 & 0.362 & 0.329\\
250 & 0 & 4.69 & 4.65 & 0.368 & 0.343\\
\hline
\end{tabular}
\label{tab:groundstateorder}
\end{table} 

\subsection{Protocol 2: Thermally quench from high $T$ to $T_f < T_{\rm cryst}$ and monitor relaxation at fixed $T_f$}
\label{subsec:protocol2}

\begin{figure}[htbp]
\includegraphics[width=3.1in]{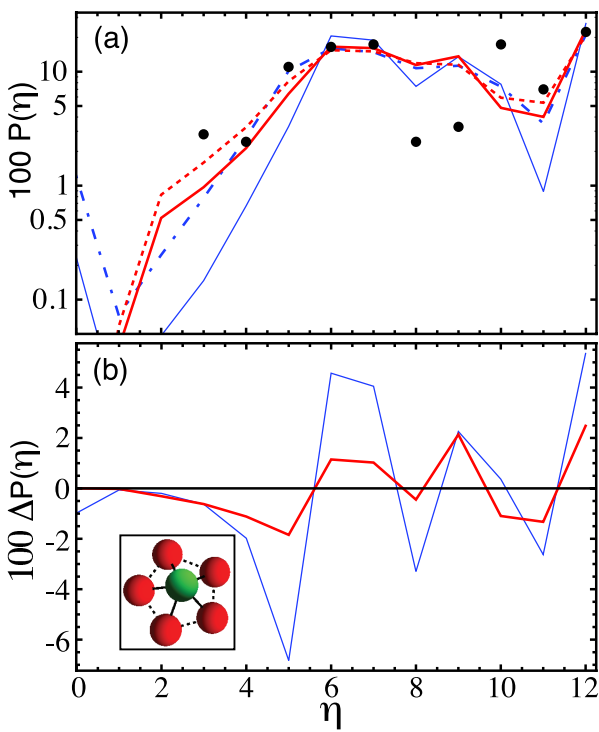}
\caption{Evolution of degree distributions $P(\eta)$.  (a) $P(\eta)$ after $|\dot{T}| = 10^{-6}$ quenches to $T_f$.  Curves show data averaged over times $0 \leq t \leq 10^{4}$ for colloids (dash-dotted) and polymers (dashed) and $0.999\cdot10^{7} \leq t \leq 10^{7}$ (colloids; light solid, polymers; heavy solid).  Solid circles show results averaged over the 6 distinct $N=100$ Barlow packings.\cite{hopkins11}  Panel (b) shows $\Delta P(\eta)$, obtained by subtracting the small-$t$ data shown in panel (a) from large-$t$ data for colloids (lighter blue line) and polymers  (heavier red line). Inset: A common five-fold symmetric structure present on the surface of nonequilibrium crystallites.  The green monomer has $\eta = 5$.}
\label{fig:degreedistrib}
\end{figure}

As shown in Table \ref{tab:groundstateorder}, polymeric and colloidal crystallites of the same $N$ have nearly equal values of both $\left<N_c/N\right>$ and $\left<N_{cp}/N\right>$ at $T_f = (7/8)T_{\rm cryst}$.
They also have similar structure (\textit{i.e.} $P(\eta)$; \textit{cf.}\ Fig.\ \ref{fig:degreedistrib}).
This helps establish that comparison of relaxation dynamics in polymeric and colloidal crystallites at fixed $T = T_f$ is a reasonable approach to isolating the role played by topology in controlling the approach of crystallites to equilibrium.

\begin{figure}[htbp]
\centering
\includegraphics[width=3.1in]{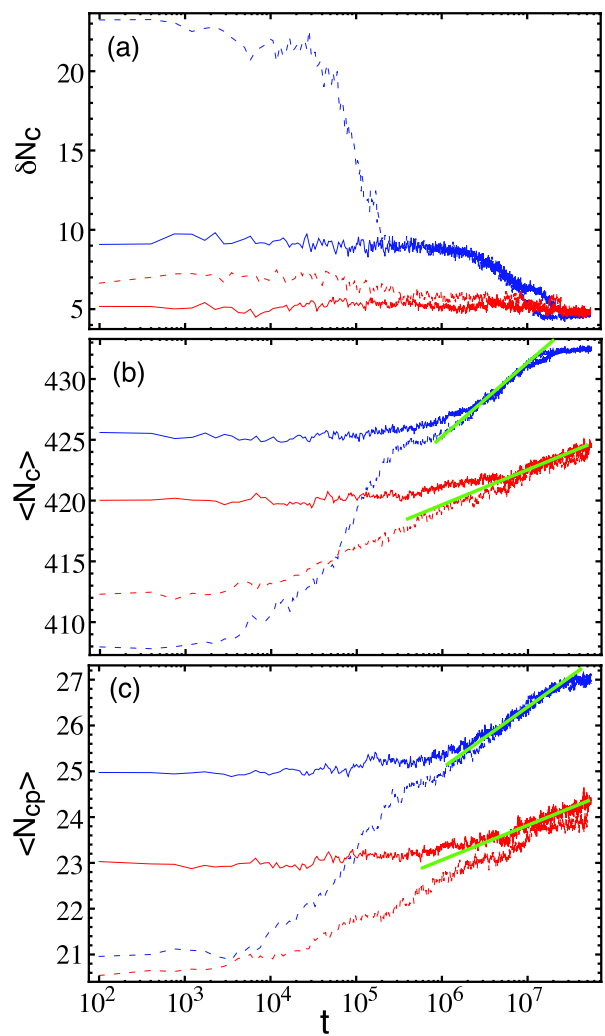}
\caption{Nonequilibrium structural relaxation for polymeric (red) and
colloidal (blue) systems after thermal quenches from $T_i > T_{\rm
cryst}$ to $T_f = (7/8)T_{\rm cryst}$ at $|\dot{T}| = 10^{-6}$ (heavier
solid lines) and $10^{-7}$ (lighter dashed lines).  Results are averaged over $104$ independent $N = 100$ samples and are plotted as a function of time $t$ following the termination of the quenches: (a) Standard deviation in the number of contacts $\delta N_c =\left<N_c^{2}\right>-\left<N_c\right>^{2}$, (b) mean number of
contacts $\left<N_c\right>$, and (c) mean size of the close-packed
cores $\left<N_{cp}\right>$.  In (b) and (c), the solid green lines show fits
to logarithmic behavior at long times (\textit{cf.}\ Table \ref{tab:Deltanalysis}); the lines are extended as a guide to the eye.}
\label{fig:m6glassy}
\end{figure}

We now examine systems quenched at different rates and monitor their evolution as a function of time $t$ following termination of the quenches to $T_f$.
Measures such as the evolution of contact order and several measures of local and nonlocal rearrangements after different waiting times $t_w$ show that polymeric crystallites possess significantly slower long-time relaxation dynamics at equal values of $T/T_{\rm cryst}$ despite their similar structure.
Studies of system size dependence show that this ``topological'' slowdown is related to the more correlated rearrangements imposed by uncrossable covalent backbones, and strengthens with increasing $N$.

{\it Contact number:} Fig.~\ref{fig:m6glassy} shows the evolution of  $\delta N_c$, $\langle N_c \rangle$, and $\langle N_{cp}\rangle$ in polymeric and colloidal systems as a function of
time $t$ following thermal quenches to $T_f$ at rates $|{\dot T}| = 10^{-6}$ and $10^{-7}$.  
$\delta N_c$ is the root-mean-square fluctuation in the number of contacts averaged over
an ensemble of collapse trajectories.  
The strong increase in $\left<N_c\right>$ and $\langle N_{cp} \rangle$ and drop in $\delta
N_c$ after $t \simeq 10^{5}$ for $|{\dot T}| = 10^{-6}$ in
Fig.~\ref{fig:m6glassy}(a) suggests that the crystal nucleation rate
for colloids is $r_{c} \approx 10^{-5}$.  In contrast, polymers do not
show a rapid increase in the number of contacts (or concomitant
decrease in the contact number fluctuations) at these thermal quench
rates, showing that the crystal nucleation rate for ($N = 100$) polymers is $r_{p}
\gtrsim 10^{-5}$.

\begin{figure}[h!]
\centering
\includegraphics[width=3.5in]{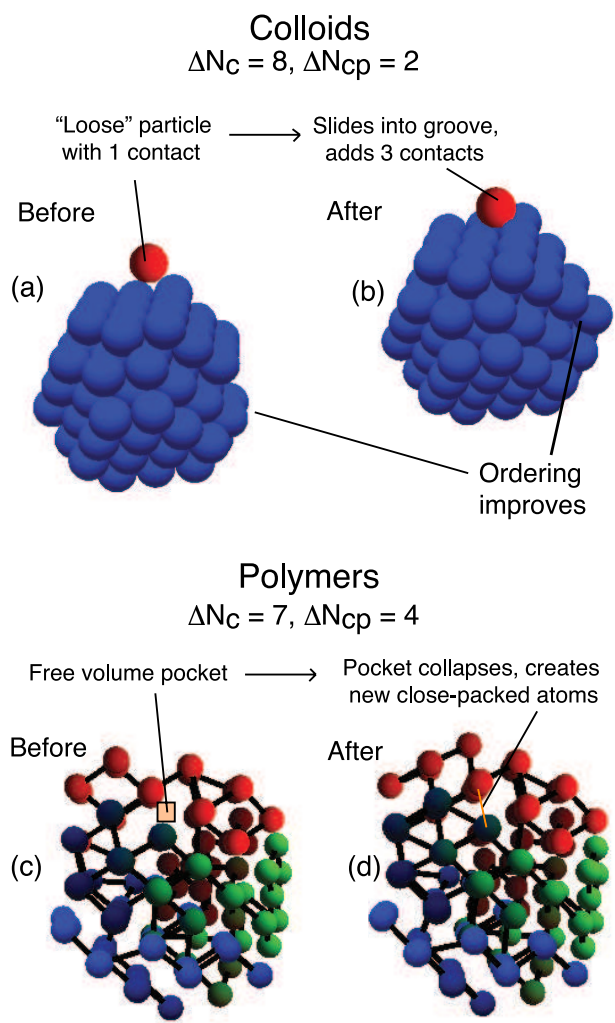}
\caption{Visualizations of typical moderately-sized rearrangement
events for polymeric and colloidal crystallites. Panels (a) and (b)
show the colloidal crystallite before and after a colloidal
rearrangement event with $\Delta N_c = 8$ and $\Delta N_{cp}=2$ and
(c) and (d) show the polymeric crystallite before and after a
polymeric rearrangement event with $\Delta N_c = 7$ and $\Delta
N_{cp}=4$.  The orange color shading in (c) and line in (d) show where new close-packed particles are added.}
\label{fig:midsizerearrnge}
\end{figure}

As shown in Fig.~\ref{fig:m6glassy} (b) and (c), at long times $t > 10^{6}$
both polymeric and colloidal systems show evidence of logarithmic
relaxation toward equilibrium. The logarithmic increase in
$\left<N_c(t)\right>$ and $\left<N_{cp}(t)\right>$ is consistent
with thermal activation over large energy barriers and transitions
between metastable, globule-like states and near-equilibrium
crystallites.\cite{socci94, taylor09}
Relaxation is also slowed by the increasing mechanical rigidity
associated with increasing $N_c$.\cite{tkachenko99} The approach to
equilibrium occurs through thermally-activated rearrangements of
particles associated with ``soft modes'' (\textit{cf.}\ Figs.\ \ref{fig:NcNcpanddeltas}-\ref{fig:midsizerearrnge}), which are known to play a significant role in structural and stress relaxation in supercooled liquids.\cite{widmer08}  
At larger $t$, the slopes (indicated by green solid lines) are clearly larger for
colloids than for polymers; values of
$\partial\left<N_c\right>/\partial \ln_{10} t$ and
$\partial\left<N_{cp}\right>/\partial \ln_{10} t$ fit over the range $10^{6.5} \leq t \leq 10^{7}$ are given in Table \ref{tab:Deltanalysis}.  

For $t > 10^{7}$, values of $\partial\langle N_c(t)\rangle/\partial\log_{10}(t)$ and $\partial\langle N_{cp}(t)\rangle/\partial\log_{10}(t)$ decrease for colloidal systems as they enter a terminal relaxation regime associated with ergodic exploration of their full free energy landscape.
The approach to the ergodic limit can be clearly seen in the vanishing of ``history'' dependence for systems quenched at different rates, \textit{i.e.}\ curves for different $\dot{T}$ overlap at large $t$. 
In contrast, for polymers, history dependence and faster logarithmic relaxation persist up to the maximum time $t = 5\cdot10^7$.
This shows that polymers possess slower characteristic relaxation rates $s_{\rm slow}$ at the same value of $T/T_{\rm cryst}$ even though the crystallites are less mechanically rigid (since they have fewer contacts and are at higher absolute $T$).

While observation of the crossover into this terminal relaxation regime for polymers with $N=100$ and $N=250$ monomers is made unfeasible by the limitations of current computer power, in this paper we are primarily concerned with the nonequilibrium dynamics of crystallization and the logarithmic ``pre-terminal'' relaxation regime of crystallite growth.
We now analyze the role of topology on relaxation dynamics within the pre-terminal regime by examining the evolution of more detailed measures of crystalline order.

{\it Degree distribution:} Figure \ref{fig:degreedistrib} shows results for the evolution of $P(\eta)$ following $|\dot{T}| = 10^{-6}$ quenches.
Polymeric and colloidal crystallites have similar degree distributions for intermediate and high $\eta$, indicating the crystallites' inner cores are similarly structured.
However, covalent backbones produce greater differences at crystallite surfaces.
Polymer topology requires $\eta \geq 2$ for chemically interior monomers and $\eta \geq 1$ for chain ends, while monomers in colloidal systems can have any degree of connectivity consistent with steric constraints (here, $0 \leq \eta \leq 12$).  
Because of this difference in connectivity, polymer crystallites include a higher fraction of monomers with $2 \leq \eta \leq 4$ (panel (a)); this difference strengthens as $t$ increases and systems approach equilibrium.

Panel (b) shows $\Delta P(\eta)= P(\eta; t = 10^7) - P(\eta; t = 0)$ to highlight the changes in $P(\eta; t)$ over an interval $\Delta t = 10^{7}$ following the end of the quench to $T_f$.  
Notable features common to colloidal and polymeric crystallites and depicted more clearly than in panel (a) include sharply negative $\Delta P(5)$ associated with the annealing out of 5-fold symmetric structures that tend to form on the surface of metastable crystallites (see inset), and sharply negative $\Delta P(11)$ associated with annealing out of defects within crystallite cores.  The latter process is closely associated with the increase in $\langle N_{cp}(t)\rangle$ shown in Fig.\ \ref{fig:m6glassy}.
However, the key result shown here is that changes in $P(\eta)$ are uniformly smaller for polymers despite the higher absolute $T$.

While the equilibrium $P(\eta)$ distribution is difficult to calculate for large $N$ without resorting to advanced Monte Carlo techniques,\cite{taylor09,seaton10} one measure of evolving crystalline order in our systems can be obtained by comparing $P(\eta; t)$ to that of a known reference system. 
Barlow packings\cite{barlow1883} are hard-sphere packings composed of layered hexagonal-close-packed planes; their three-dimensional order may be fcc, hcp, or mixed fcc/hcp, but they possess no defects (\textit{e.g.}\ stack faults.) 
These are ``reference'' low energy structures for our model in the limit of steep hard-core repulsions and short-range attractions.
The solid circles in Fig.\ \ref{fig:degreedistrib}(a) show $P(\eta)$ averaged over the six $N = 100$ Barlow packings.\cite{hopkins11}

Our simulation data show that crystallites at the end of the pre-terminal relaxation regime possess Barlow-like order.
Interestingly, $\langle N_c \rangle$ is slightly higher (Fig.\ \ref{fig:m6glassy}) in the simulated systems than in the Barlow packings ($\langle N_c^{\rm Barlow} \rangle = 421$), while values of $P(12) = \langle N_{cp}/N \rangle$ are similar.
These effects are attributable to finite temperature, stiffness of core repulsions and range of attractive interactions.
Other differences between our systems and Barlow packings are attributable to small deviations from equilibrium, the fact that our crystallites may be stack-faulted, and (for polymers) entropic factors such as blocking.\cite{hoy10}
That the above complexity can be understood by examining $N_c$, $N_{cp}$, and $P(\eta)$ shows that this combination of metrics constitutes an effective ``crystal-agnostic''\cite{tenwolde96} measure of order.\footnote{Indeed, previous coarse-grained Monte Carlo studies of polymer crystallization\cite{karayiannis09pre,karayiannis09prl,karayiannis09jcp} focusing on dense bulk systems have employed similar structural measures.}

In the above protocol 2 subsections, we have focused on results for $N = 100$.  
System size effects are minor, \textit{e.g.} slower logarithmic growth of crystalline order with increasing $N$ and shift of $P(\eta)$ towards higher $\eta$.
Examining rearrangements within crystallites provides additional insight into $N$- and topology-dependent effects on the glassy dynamics of crystallite formation and is discussed in the following subsections.

{\it Adjacency matrix autocorrelation function:} Next we examine the
decorrelation of contacts between neighboring particles during the
approach to equilibrium at $T_f = (7/8)T_{\rm cryst}$.  
We first examine effects of quench rate and waiting time on $N=100$ systems, and then examine $N$-dependence for evolution following slow quenches.
The adjacency matrix autocorrelation function $P_{\rm AMAC}(t_w,t')$ displays several
important features (Fig.~\ref{fig:PAMAC}):

\begin{figure}[h!]
\centering
\includegraphics[width=3.1in]{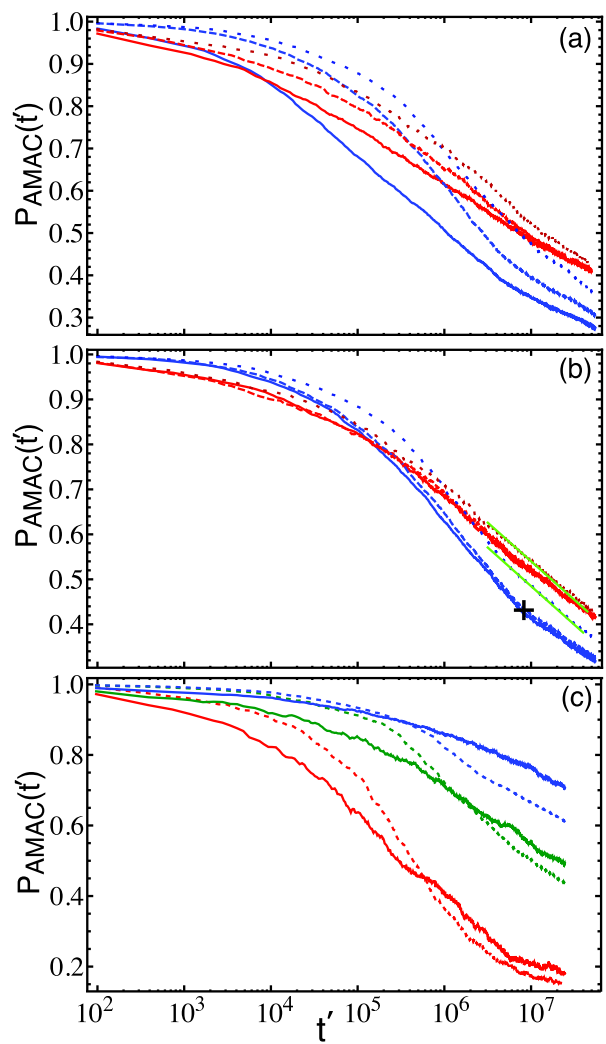}
\caption{The adjacency matrix autocorrelation function $P_{AMAC}(t_w,t')$ following (a) $|\dot{T}| = 10^{-6}$ and (b)
$10^{-7}$ thermal quenches.  Protocols and systems for panels (a-b) are the same as in Fig.\ \ref{fig:m6glassy}. Results for polymers (colloids) are shown in red (blue).
$P_{AMAC}(t_w,t')$ is measured as a function of time $t'$ after waiting times $t_w=0$,
$t_w = 10^{5.5}$, and $t_w = 10^{6.5}$ (solid, dashed, and dotted curves, respectively).  The solid green
lines in (b) show fits to logarithmic behavior at long times for $t_w=10^{6.5}$ and the black (+) sign denotes crossover into the terminal relaxation regime for colloids.  Panel (c) shows how differences in $P_{AMAC}(t_w = 0; t')$ between slowly quenched ($|\dot{T}| = 2.5\cdot 10^{-8}$) polymeric and colloidal systems vary with $N$.  Solid curves represent polymeric and dashed curves represent colloidal results, while colors indicate $N=40$ (red), $N = 100$ (green), and $N=250$ (blue).  Each curve in (c) represents an ensemble average over 40 statistically independent systems.}
\label{fig:PAMAC}
\end{figure}

\textbf{(i)}:\ Polymeric rearrangement events are more frequent at low
$t'$ because of the higher absolute $T$.  However, rearrangements are
slower at large $t'$ despite the higher absolute $T$.  The slower
decay arises from the covalent bonds in polymers that restrict the
motion of monomer $i$ to the plane tangent to the vector
$\vec{r}_{i+1}-\vec{r}_{i-1}$.  Although contributions from permanent
covalent bond contacts are excluded from the definition of $P_{\rm
AMAC}(t_w,t')$, in compact crystallites connectivity to chemically
distant monomers produces long-range suppression of
contact-breaking.\footnote{This could be examined quantitatively by
excluding successively more distant chemical neighbors  (\textit{e.g.}\ 2nd nearest, 3rd nearest),
and by considering only chemically interior sections of polymers.
A detailed analysis is left for future studies of near-equilibrium and
equilibrium systems.}

\textbf{(ii)}:\ Following the $|\dot{T}| = 10^{-6}$ thermal quenches,
$P_{\rm AMAC}(t_w,t')$ for both polymers and colloids display strong
$t_w$-dependence as shown in Fig.~\ref{fig:PAMAC}(a).  We find an
increase in the length and height of the low-$t'$ ``plateau'' near
$P_{\rm AMAC}(t_w,t') \approx 0.95$ with increasing $t_w$, 
similar to the behavior of the plateau in $S(q,t_w,t')$ during the
aging process in structural glasses.\cite{kob00}  Aging effects are
stronger for colloids than for polymers because colloids are further
from equilibrium at the termination of the quenches (Fig.\ \ref{fig:m6glassy}(a)).

\textbf{(iii)}: For slower thermal quenches ($|\dot{T}| = 10^{-7}$;
Fig.\ \ref{fig:PAMAC} (b)), similar but much weaker aging effects are
present.  
Results for $t_w=0$ and $t_w = 10^{5.5}$ are indistinguishable to within statistical uncertainties for both polymers and colloids. 
Aging is delayed in part because the additional time to quench from $T_{\rm cryst}$ to $T_{f}$ provided by the factor of 10 decrease in quench rate is larger than $r_{c}^{-1} \sim 10^{5}$ and $r_{p}^{-1} > 10^{5}$ (Fig.\ \ref{fig:m6glassy}(a)), and in part because for $t_w \gtrsim 10^{6.5}$ systems have crossed into the preterminal (logarithmic) relaxation regime at $t' = t-t_w = 0$.
Both polymer and colloidal crystallites continue to slowly
add contacts and close-packed monomers, and the $t_w$-dependence should not vanish
until equilibrium values of $\left<N_c\right>$ and
$\left<N_{cp}\right>$ are reached.

\textbf{(iv)}:\ At large $t'$, the adjacency matrix autocorrelation
function decays logarithmically.
The crossover to logarithmic decay of $P_{\rm AMAC}(t_w,t')$ corresponds to the pre-terminal regime
of logarithmic growth of $\left<N_c\right>$ and $\left<N_{cp}\right>$
(Fig.\ \ref{fig:m6glassy}(b) and (c)).  
For colloids and small $t_w$, a decrease in the slope of this logarithmic decay corresponding to crossover into the terminal relaxation regime is indicated by the $+$ symbol in Fig.\ \ref{fig:PAMAC}(b).
No such change in slope occurs for polymers over the same range of $t'$.
This is consistent with the idea that local relaxations in polymers are slower due to chain-connectivity constraints on rearrangements.

\textbf{(v)}:\ Figure \ref{fig:PAMAC}(c) illustrates the variation of $P_{\rm AMAC}(t_w=0; t')$ with $N$ for slowly quenched systems.  
Characteristic contact decorrelation rates $s_{cont}$ decrease sharply with increasing $N$; for example, the low-$t'$ plateau lengthens with increasing $N$, and the $t'$ at which $P_{\rm AMAC} = 0.8$ is 2-3 orders of magnitude greater for $N = 250$ than for $N = 40$. 
Since this $N$-dependent decrease in $s_{cont}$ is similar for polymeric and colloidal crystallites (which, as we have shown, possess similar structure), we attribute it to the increasing contribution of crystallite cores where reneighboring dynamics are slow because $\eta$ is high and particles are more sterically constrained.
Similarities and differences between polymeric and colloidal results are consistent with those expected from \textbf{(i)}.
For all $N$, as in panels (a-b), polymers relax faster than colloids at low $t'$ because $T_f = (7/8)T_{\rm cryst}$ is higher, and slower at large $t'$ because of topologically restricted rearrangement (\textit{cf.} Figs.\ \ref{fig:NcNcpanddeltas}-\ref{fig:midsizerearrnge}.)
Both the ``crossover'' $t'$ at which $P_{\rm AMAC}^{coll} = P_{\rm AMAC}^{pol}$ and the ratio $P_{\rm AMAC}^{pol}/P_{\rm AMAC}^{pol}$ beyond this crossover time increase with increasing $N$.

In this paper we focus on the glassy dynamics of crystallite
\textit{formation} (where about half of the contacts existing at the termination of the quench have not been broken), not complete reorganization.  Below, we show that there are
significant differences between large-scale polymeric and colloidal
rearrangements in this regime.  
In the remainder of this section, we will focus on $N=100$ collapsed states generated using protocol $2$ with thermal quench rate $|\dot{T}|=10^{-7}$.

{\it Statistical comparison of rearrangements in polymeric and
colloidal crystallites:} We describe rearrangement events using the
two-dimensional parameter space $(N_c, N_{cp})$, where
tangent-sticky-sphere polymers and colloids have the same inherent
structures.\cite{stillinger95} In Fig.\ \ref{fig:NcNcpanddeltas} we
show the probability distribution $P(\Delta N_c, \Delta N_{cp})$ for crystallite rearrangements to
add $\Delta N_c$ contacts and $\Delta N_{cp}$ close-packed
particles in crystallites over time intervals $\Delta t =10^{3}$.\footnote{Note that for the physically reasonable
values $m=10^{-24}\rm kg$, $D=1\rm nm$, and $\epsilon \simeq 10k_BT$
at room temperature, $\tau \simeq 5\rm ps$, and timescales $\sim \Delta t$ can be probed by neutron spin echo experiments, \textit{e.g.}\ for the purpose of characterizing protein dynamics.\cite{biehl11}}
Results are calculated for the range $10^{6.5} \leq t \leq 10^7$ where both colloids and polymers are in the preterminal relaxation regime.  $P(\Delta N_c, \Delta N_{cp})$ is proportional to
the integrated rate
\begin{equation}
R(\Delta N_c, \Delta N_{cp}) = \int \int s(\Delta N_c, \Delta N_{cp}, N_c^{0}, N_{cp}^{0}) dN_c^{0} dN_{cp}^{0}
\label{eq:RP}
\end{equation}
for all transitions that add $\Delta N_c$ contacts and $\Delta
N_{cp}$ close-packed particles in crystallites originally posessing $N_c^{0}$
contacts and $N_{cp}^{0}$ close-packed particles, {\it i.e.} $R(\Delta
N_c, \Delta N_{cp}) \simeq P(\Delta N_c, \Delta N_{cp})/\Delta t$.  
This becomes exact in the limit $R(\Delta N_c, \Delta N_{cp})\Delta t \to 0$.
Thus, the data in Fig.\ \ref{fig:NcNcpanddeltas} provides a basis for
comparing free energy barriers and transition rates in these systems.

\begin{figure*}[htbp]
\centering
\includegraphics[width=6.85in]{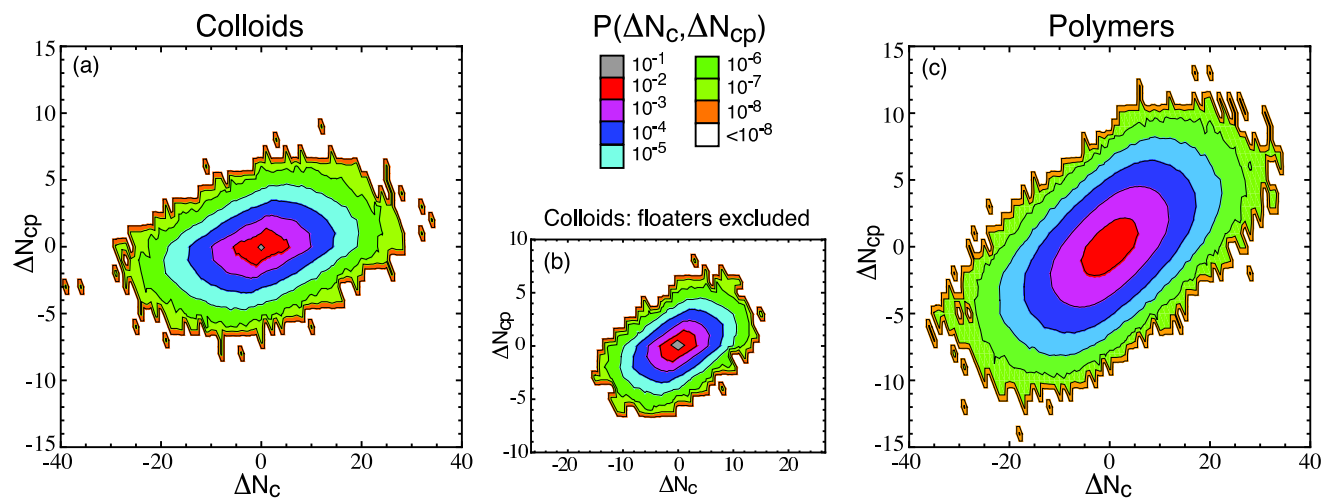} 
\caption{Probability distributions $P(\Delta N_c, \Delta N_{cp})$ for
changes in the number of contacts and number of close-packed monomers
over time intervals of $\Delta t = 10^{3}$ following $|\dot{T}| =
10^{-7}$ thermal quenches from $T_i > T_{\rm cryst}$ to $T_f/T_{\rm
cryst} = 7/8$ for (a) colloids, (b) colloids with ``floaters''
excluded, and (c) polymers.  The different colored regions indicate
bins in probability that differ by factors of $10$.  Results are for
the same systems analyzed in Figs.\ \ref{fig:m6glassy} and
\ref{fig:PAMAC}(b).  To capture the ``pre-terminal'' logarithmic
relaxation regime, only results for $10^{6.5} \leq t \leq 10^7$ are presented.}
\label{fig:NcNcpanddeltas}
\end{figure*}

Figure~\ref{fig:NcNcpanddeltas} and Table~\ref{tab:Deltanalysis}, which 
characterize the shape of the distribution $P(\Delta N_c, \Delta N_{cp})$,
illustrate the dramatic differences between polymeric and colloidal
rearrangements during logarithmic relaxation.  Polymeric
rearrangements are significantly more correlated than those for
colloids. For example, the correlation coefficient
\begin{equation}
c = \displaystyle\frac{\big<\left(\Delta N_c - \left<\Delta N_c\right>\right)\left(\Delta N_{cp} - \left<\Delta N_{cp}\right>\right)\big>}{\sqrt{\big<\left(\Delta N_c - \left<\Delta N_c\right>\right)^2 \left(\Delta N_{cp} - \left<\Delta N_{cp}\right>\right)^2\big>}},
\label{eq:samplec}
\end{equation}
where the averages are taken over all rearrangements, is larger by a
factor of $2.5$ and the cross correlation $\left<\Delta N_c \Delta
N_{cp}\right>$ is larger by a factor of $7.4$ for polymers compared to 
colloids.  The second moments of $P(\Delta N_c, \Delta
N_{cp})$, $\left<(\Delta N_c)^2\right>$ and $\left<(\Delta
N_{cp})^2\right>$, are also larger for polymers than colloids.

\begin{table}[h]
\centering
\small\caption{Statistical analysis of the data presented in Fig.
\ref{fig:NcNcpanddeltas}.  The top two rows are calculated by fitting
to data in Fig.\ \ref{fig:m6glassy} in the preterminal regime ($10^{6.5} \leq t \leq 10^{7}$)  The middle column shows data for colloidal rearrangements excluding ``floaters''.}
\begin{tabular}{lccc}
\hline
Quantity & Colloids & Colloids (NF) & Polymers\\
\hline
$\partial \left<N_c\right>/\partial \log_{10} t$ & 6.3 & 6.3 & 1.4 \\
$\partial \left<N_{cp}\right>/\partial \log_{10} t$ & 2.1 & 2.1 & 0.7\\
$\left<\Delta N_c^{2}\right>$ & 9.1 & 2.2 & 23\\
$\left<\Delta N_{cp}^{2}\right>$ & 0.41 & 0.32 & 3.2\\
$\left<\Delta N_c \Delta N_{cp}\right>$ & 0.66 & 0.39 & 4.9\\
$c$ & 0.23 & 0.34 & 0.58\\
\hline
\end{tabular}
\label{tab:Deltanalysis}
\end{table} 

An interesting feature of $P(\Delta N_c, \Delta N_{cp})$ is that it is
nonzero in quadrants II and IV (\textit{i. e.}\ negative $\Delta N_c \Delta N_{cp}$).
Rearrangements occur in which the number of close-packed monomers
decreases but the number of contacts increases overall, and vice versa.
These are less likely for polymers than for colloids since covalent bonds impart greater cooperativity to rearrangements in polymers.  $P(\Delta N_c, \Delta
N_{cp})$ is reasonably well fit by a 2D Gaussian functional form
\begin{equation}
P^*(\Delta N_c,\Delta N_{cp}) \propto e^{-E(\Delta N_c)^2 + F \Delta
N_c \Delta N_{cp} - G(\Delta N_{cp})^2},
\label{eq:Pstar}
\end{equation}
where $E$ is similar for polymers and colloids, $F$ is larger for
polymers, $G$ is larger for colloids, and $-F/(2EG)$ is larger for
polymers.  These results are consistent with our finding that $c$ is
larger for polymers than for colloids.  

The fit $P^*(\Delta N_c,\Delta N_{cp})$ does, however, fail to capture several key features of 
rearrangements.  In particular, $P(\Delta N_c, \Delta N_{cp})$ has a
sharper peak at the origin and ``fat'' (wider than Gaussian) tails.
The fat tails of $P(\Delta N_c, \Delta N_{cp})$ are dominated by
events where a monomer escapes from or rejoins the crystallite (for
colloids), or a chain end becomes ``floppy'' (for polymers). These
events tend to be associated with rearrangements with large $\Delta
N_c$ and small $\Delta N_{cp}$ for colloids, and large $\Delta N_c$
and $\Delta N_{cp}$ for polymers (the difference being that covalent
bonds impart a large degree of coupling between the exterior and
close-packed interior of crystallites).  While rare, these events may
dominate large rearrangements in which many changes of
contacts occur.

To quantify this effect, we analyze colloidal rearrangements with
``floaters'' excluded, {\it i.e.} where all particles possess at least
one interparticle contact both before and after the rearrangement.
Since monomers in polymers always have at least one contact, excluding
floaters provides a more direct method to quantify the role of topology
in controlling the rearrangements occurring during slow crystallite growth.
Figure \ref{fig:NcNcpanddeltas}(b) and the ``Colloids (NF)'' data in Table
\ref{tab:Deltanalysis} present this analysis.  The ``no-floater''
subset of colloidal rearrangements possesses significantly smaller
fluctuations (in terms of $\left<\Delta N_c^{2}\right>$ and
$\left<\Delta N_{cp}^{2}\right>$) and is more correlated (in terms of
$\left<\Delta N_c \Delta N_{cp}\right>$).  However, no-floater
colloidal rearrangements remain significantly less correlated than
those for polymers.
Thus we claim that the distributions $P(\Delta N_c, \Delta N_{cp})$ differ dramatically for
polymers and colloids ({\it i.e.} are more correlated for polymers) precisely
because their free energy landscapes are different, which in turn is a
direct consequence of the presence or lack of covalent bonds.  

The large width of the distribution $P(\Delta N_c, \Delta N_{cp}$) implies that the dynamics of crystallite growth in the pre-terminal relaxation regime are highly
heterogeneous, especially for polymers.  One source of dynamical heterogeneity, as suggested by
Fig.\ \ref{fig:NcNcpanddeltas}, is surface effects.
Rearrangements at the surfaces of crystallites (where monomers possess low $\eta$) are qualitatively different than those occurring within the close-packed cores.  
To illustrate the role of surface effects and covalent bonds, we now compare typical rearrangement events characteristic of those occurring in ($N=100$) polymeric and colloidal crystallites.

{\it Visualization of typical rearrangements in polymeric and
colloidal crystallites:} Figure \ref{fig:midsizerearrnge} shows a
colloidal rearrangement with $\Delta N_c = 8$, $\Delta N_{cp} = 2$
((a) and (b)) and a polymeric rearrangement with $\Delta N_c = 7$,
$\Delta N_{cp} = 4$ ((c) and (d)).  These are roughly equiprobable,
with $\log_{10} P(\Delta N_c,\Delta N_{cp}) \approx -3.5$, and the
ratio $\Delta N_{cp}/\Delta N_{c} \approx c$. (See Table
\ref{tab:Deltanalysis}.)  Before the colloidal rearrangement, the red
monomer has only one contact; during the rearrangement, it settles
into a groove on the surface of the crystallite and adds three new
contacts.  At the same time, the right hand side of the crystallite
undergoes a large rearrangement, which improves its stacking order,
and five other contacts and two other close-packed atoms are added
elsewhere.  It is likely that the initial ``looseness'' of the red
monomer destabilizes the crystallite and gives rise to the soft mode that causes this large
rearrangement.

For polymers, the picture is \textit{qualitatively} similar, but is further complicated by the fact that the interior and exterior of crystallites are topologically connected.
Large rearrangements are often initiated when a chain end is at the surface
and relatively loose.  Note that in Fig.~\ref{fig:midsizerearrnge} (c)
and (d) the color varies from red to green to blue as the monomer
index increases from $1$ to $100$.  Before the rearrangement, the
chain end including monomer $100$ is relatively loose (with only $3$
contacts), and a pocket exists in the crystallite with above average
free volume.  During the rearrangement a segment of the
polymer including this end executes a ``flip'' that collapses the
pocket.  The chain end monomer adds three contacts, and
four other contacts and close-packed atoms are added elsewhere.

Covalent bonds suppress large rearrangements less when chain ends
exist on the exterior of crystallites.  In the rearrangement event
depicted in Fig.~\ref{fig:midsizerearrnge} (c) and (d), the path of
the covalent backbone through the crystallite is not particularly
tortuous---it proceeds in a relatively orderly fashion from
upper-right to lower-left.  Entropic factors such as ``blocking''
suppress the probability for chain ends to exist in the interior,\cite{hoy10,karayiannis09pre} otherwise large rearrangements would be
even further suppressed.

In summary, despite qualitative similarities, polymer topology
produces the large quantitative differences in slow crystallite growth and rearrangements illustrated above in Figs.\ \ref{fig:m6glassy}-\ref{fig:midsizerearrnge} and Table \ref{tab:Deltanalysis}.  Even larger and rarer polymeric rearrangements than those depicted in Fig.\
\ref{fig:midsizerearrnge} involve cooperative rearrangements of
sub-chains that do not include chain ends.  Such large-scale
rearrangements occur within the interior of the crystallite; their
initiation \textit{requires} a large (negative) $\Delta N_{c}$ and
$\Delta N_{cp}$, and hence they possess large activation energies.
These are the slowest relaxation mechanisms, and it is likely that
they control the approach to the ergodic limit in polymer
crystallites.

\section{Discussion and Conclusions}
\label{sec:discussion}

In this article, we compared the crystallization dynamics of single-chain polymers and
colloids.
The use of model systems with hard-core-like repulsive and short-range attractive interparticle
potentials yielded contact-dominated crystallization and allowed us to characterize crystalline order via measures such as the number of contacts $N_c$, the number of close-packed particles $N_{cp}$, and the contact degree distribution $P(\eta)$.
Our use of a model in which covalent and noncovalent bonds have the same equilibrium bond length yielded the same low energy structures for polymeric and colloidal systems, allowing us to isolate the role of chain topology on the dynamics of crystallite formation and growth.

Particular attention was paid to the effect of thermal quench rate.
Slow thermal quench rates yield first-order like transitions to crystallites at $T=T_{\rm cryst}$.
The ratio of $T_{\rm cryst}$ for polymers to that for colloids can be
obtained roughly by counting degrees of freedom. 
Comparison of polymeric and colloidal crystals at equal values of $T/T_{\rm cryst}$ showed they possess similar structure (\textit{i.e.} $N_c$, $N_{cp}$, and $P(\eta)$), and thus occupy similar positions on their respective free energy landscapes, despite significantly different absolute $T$. 
Higher quench rates yielded rate-dependent effects and glassy relaxation
from partially disordered to more ordered configurations.  
While the marked slowdown in dynamics at $T = T_{\rm cryst}$ and consequent rate-dependent glassy behavior for crystal-forming systems possessing phase diagrams like Fig.\ \ref{fig:one}(a) 
is understood in terms of a crossover to potential energy landscape dominated dynamics \cite{schroder00} with decreasing temperature, the role of covalent backbones (and consequently, different energy landcapes) in producing the strongly differing nonequilibrium responses for polymers and colloids reported in this paper has not been previously isolated.
We showed that although polymer crystallites nucleate faster because of the cooperative
dynamics imparted by their covalent backbones, chain connectivity
slows their relaxation towards maximally ordered structures.

Crystallites can rearrange in many different ways ({\it e.g.} with different changes $\Delta N_c$ and $\Delta N_{cp}$ in the number of contacts and close-packed monomers, respectively).  
By measuring the transition probability $P(\Delta N_c,\Delta N_{cp})$ in the regime where the degree of crystalline order exhibits slow logarithmic growth, we characterized how the rare collective rearrangement events which control the slow approach of crystallites to equilibrium are affected by polymer topology.
Significant differences between $P(\Delta N_c, \Delta N_{cp})$ in polymeric and colloidal crystallite formation are attributable to the increased cooperativity of rearrangements required by the covalent backbone.  

Strong finite size effects have been observed in both equilibrium and
nonequilibrium polymer-collapse studies.\cite{seaton10,dalnoki,lopatina11} 
Here we examined system size effects using measures of order such as $N_c$ and $N_{cp}$ that vary rapidly with the number of particles $N$ due to concomitant variation in the ratio of crystallite surface area to volume.
This variation did not change any of the qualitative features reported above, and quantitative differences were as expected; dynamical slowdown of relaxation associated with restricted motion imposed by the covalent backbone strengthens with increasing $N$.

It is well known that bond-angle interactions play a significant role in controlling crystallization of most synthetic polymers.
While this study considered flexible chains, it serves as a basis for future studies of more realistic models by elucidating the role polymer topology plays in controlling the glassy dynamics of crystallization.
Our results may also be directly applicable to understanding the collapse behavior of flexible ``colloidal polymers'', which have recently been shown to self-assemble into tunable, compact nanostructures,\cite{sacanna10} as well as very flexible natural polymers such as single stranded DNA.\cite{latinwo11} 
Future work will examine how crystallization and packing are affected by semiflexibility, as well as effects of topology on the dynamics of equilibrium crystallites.

\section{Acknowledegements}

All MD simulations were performed using the LAMMPS molecular dynamics
simulation software.\cite{plimpton95}  We thank Steven J.\ Plimpton
for developing enhancements to LAMMPS for the long-time runs, S.\
S.\ Ashwin and K.\ Dalnoki-Veress for helpful discussions, and Adam Hopkins for providing the $N = 100$ Barlow packings.\cite{hopkins11}
Support from NSF Award No.\
DMR-1006537 is gratefully acknowledged.  This work also benefited
from the facilities and staff of the Yale University Faculty of Arts
and Sciences High Performance Computing Center and NSF grant
No.\ CNS-0821132 that partially funded acquisition of the computational
facilities.

\providecommand*{\mcitethebibliography}{\thebibliography}
\csname @ifundefined\endcsname{endmcitethebibliography}
{\let\endmcitethebibliography\endthebibliography}{}


\begin{mcitethebibliography}{46}
\bibitem[foo()]{footprevcollpol}
Previous studies\cite{schnabel09} comparing collapsed states of polymers and colloids with system sizes comparable to those considered in this paper have employed much softer (Lennard-Jones) interactions for which covalent bond stretching can produce greater differences in the low energy states, and have focused on equilibrium properties.\relax
\mciteBstWouldAddEndPunctfalse
\mciteSetBstMidEndSepPunct{\mcitedefaultmidpunct}
{}{\mcitedefaultseppunct}\relax
\bibitem[foo()]{footnobondcross}
Studies examining the effects of polymer topology must not allow crossing of
  covalent bonds. The lowest barrier $\Delta U_{\rm cross}$ for covalent bond
  crossing can be estimated\cite{sukumaran05} by placing two bonds on opposite
  corners of a square. For the harmonic spring potential (Eq.\ \ref{eq:perturbed}), $\Delta
  U_{cross} = k(6 - 4\sqrt{2})$. For the temperatures and spring constant $k =
  1600\epsilon$ considered here, $\Delta U_{\rm cross} \gtrsim 10^{3} k_B T$.\relax
\mciteBstWouldAddEndPunctfalse
\mciteSetBstMidEndSepPunct{\mcitedefaultmidpunct}
{}{\mcitedefaultseppunct}\relax
\EndOfBibitem
\bibitem[foo()]{footnoicos}
The small $r_c$ suppresses local icosohedral order in favor of local fcc/hcp
  order; this is likely one of the reasons these systems exhibit direct
  ``gas''-solid transitions.\relax
\mciteBstWouldAddEndPunctfalse
\mciteSetBstMidEndSepPunct{\mcitedefaultmidpunct}
{}{\mcitedefaultseppunct}\relax
\EndOfBibitem
\bibitem[foo()]{footstickysphere}
Results in the sticky-sphere ($k \to \infty$) limit are known to provide a
  reference for perturbative studies of other forms of $U(r)$ with
  hard-core-like repulsions and differently ranged
  attractions.\cite{cochran06,foffi00} Eq.\ \ref{eq:perturbed} is also a
  first-order approximation for the depletion interactions typical of
  nano-self-assembling colloidal and ``colloid-polymer'' systems.
  \cite{meng10,sacanna10}\relax
\mciteBstWouldAddEndPunctfalse
\mciteSetBstMidEndSepPunct{\mcitedefaultmidpunct}
{}{\mcitedefaultseppunct}\relax
\EndOfBibitem
\bibitem[foo()]{foot12}
The curvature in $E_{pot}(T)$ for polymers indicates the well understood
  progression from extremely good solvent behavior at high $T$ to ``poor''
  (but, for this potential, not poor enough to induce collapse into globules)
  solvent behavior at $T$ just above $T_{\rm cryst}$.\relax
\mciteBstWouldAddEndPunctfalse
\mciteSetBstMidEndSepPunct{\mcitedefaultmidpunct}
{}{\mcitedefaultseppunct}\relax
\EndOfBibitem
\end{mcitethebibliography}

\end{document}